\pdfoutput=1
\documentclass[twocolumn,twoside,slac_two]{revtex4_mine}
\usepackage{graphicx}
\usepackage{fancyhdr}
\begin{document}

\title{Search for Extra Dimensions in the Diphoton Channel}


\author{Selda Esen, and CMS Collaboration}
\affiliation{Department of Physics, Brown University, Providence, RI, USA}

\begin{abstract}

We present a search for extra dimensions in the diphoton channel using the CMS detector at the Large Hadron Collider. The search is focused on the forthcoming 2009--2010 run at 10 TeV center$-$of$-$mass energy and $\sim 100$ pb$^{-1}$ of data. We discuss event selection and optimization, as well as data-driven methods of estimating various backgrounds and efficiencies. The dominant source of background after all the selection requirements is SM diphoton production. We quote the sensitivity of the search both in terms of limits on the parameters of large and warped extra dimensions in the case of no excess observed, and in terms of signal discovery significance, if an excess is seen in data.

\end{abstract}

\maketitle

\thispagestyle{fancy}


%%%%%%%%%%%%%%%%%%%%%%%%%%%%%%%%%%
\section{Introduction}

New approaches which exploit the geometry of extra spatial dimensions have been  proposed  to resolve the hierarchy problem of the Standard Model \cite{add,rs,intro}.   The hierarchy problem refers to the large difference between the Planck scale ($M_{Pl}\sim10^{16}$ TeV) where gravity is expected to be strong, and the scale of electroweak symmetry breaking ($\sim1$ TeV).
%There are a set of various scenarios \cite{add,rs} where 
The scenario proposed by Arkani-Hamed, Dimopoulos and Dvali (ADD) 
deals with the case in which the hierarchy is generated by a large volume  
for the extra dimensions \cite{add}. The Randall-Sundrum (RS) scenario suggests that the observed hierarchy is created by a warped extra dimension~\cite{rs}.
%an exponential warp factor arising from a 5-dimensional non-factorizable geometry \cite{rs}. 
%
If these theories, in fact, describe the origin of the observed hierarchy, then their signatures
should appear at TeV scale experiments.

Here we present a search for large extra dimensions and warped extra dimensions in the diphoton channel using the CMS detector at the Large Hadron Collider (LHC).The search is focused on the forthcoming 2009--2010 run at 10 TeV center-of-mass energy ~\cite{led_CMS, RS_CMS}. 

%%%%%%%%%%%%%%%%%%%%%%%%%%%%%%%%%%
\section{Large Extra Dimensions}

Compact large Extra Dimensions (ED) are an intriguing proposed solution of the hierarchy problem of the Standard Model (SM), which refers to the puzzling fact that the fundamental scale of gravity, $M_{\rm Pl}\sim 10^{19}$~GeV, is so much higher than the electroweak scale $M_{\rm EWSB} \sim 10^3$~GeV.  With such a difference in scale, it is difficult to protect the Higgs mass from radiative corrections without a very high degree of fine-tuning.

The original proposal to use ED to solve the hierarchy problem was presented by Arkani-Hamed, Dimopoulos, and Dvali (ADD) \cite{add}.  They posit a scenario whereby the SM is constrained to the common $3+1$ space-time dimensions, while gravity is free to propagate through the entire multidimensional space (``bulk'').  Because of this, the gravitational force is effectively diluted, having undergone a Gauss's Law reduction in the flux.  The fundamental Planck scale, $M_D$ is therefore related to the apparent scale, $M_{\rm Pl}$, according to the formula:
\begin{equation}
M_D^{(n_{\rm ED}+2)} \sim \frac{M_{\rm Pl}^2}{R^{n_{\rm ED}}},
\end{equation}
where $R$ and $n_{\rm ED}$ are the size and number of the ED, respectively.  Current experimental constraints allow a scenario with $n_{\rm ED}\ge 2$ corresponding to ED sizes $\lesssim 10^{-1}$~mm.

For virtual graviton processes, the effects of ED are parameterized via a single variable $\eta_G = {\cal F}/M_S^4$, where ${\cal F}$ is a dimensionless parameter of order unity\footnote{Here $M_S$ is a ultraviolet cutoff used to regularize the calculation; it is expected to be close to $M_D$, but may be somewhat different from the latter, thus emphasizing complementary nature of virtual graviton exchange and other ways of probing large ED, e.g. by exploring direct production of gravitons in association with a photon or a jet.}. Several conventions for ${\cal F}$ are used in the literature~\cite{GRW,Hewett,HLZ}; we use the one after Han, Lykken, and Zhang ~\cite{HLZ}:
\begin{equation}
        {\cal F} = \left\{ \begin{array}{ll} 
         \log\left( \frac{M_S^2}{\hat s} \right), & n = 2 \\
           \frac{2}{n-2}, & n > 2,
           \end{array} \right. \label{eq:HLZ}
\end{equation}
which is the only convention to date that has an explicit dependence on the number of ED.

Collider phenomenology of models with large ED has been studied in many details~\cite{GRW,HLZ,Peskin,Hewett}. In this analysis we present an early measurement of virtual graviton production in the diphoton final state with early data of 10~TeV $pp$ collisions at CMS ~\cite{led_CMS}.

%%%%%%%%%%%%%%%%%%%%%%%%%%%%%%%%%%
\subsection{Photon Identity at CMS and Event Selection}

Due to the large amount of material in the CMS tracking detectors, electrons and photons in CMS tend to radiate and convert. The resulting pattern of energy in the electromagnetic calorimeter (ECAL) looks like several closely-spaced clusters of energy (``supercluster''). 
We reconstruct photons more than 50 GeV of $E_{T}$ in the barrel region ($|\eta|<1.5$) by considering ECAL superclusters that do not have associated pixel detector hits (which would indicate an electron or positron) and have a loose requirement on the maximum amount of hadronic energy near the supercluster. We can reduce jets misidentified as photons significantly without a large loss in efficiency by placing requirements on a set of isolation variables. These variables are \textit{HadronicOverEM},\textit{Tracking Isolation}, \textit{ECAL Isolation}, and \textit{HCAL Isolation} (see details in ~\cite{led_CMS}). 

\begin{figure}[tb]
%\begin{figure*}[floatfix]
\centering
\includegraphics[width=80 mm,height=40 mm]{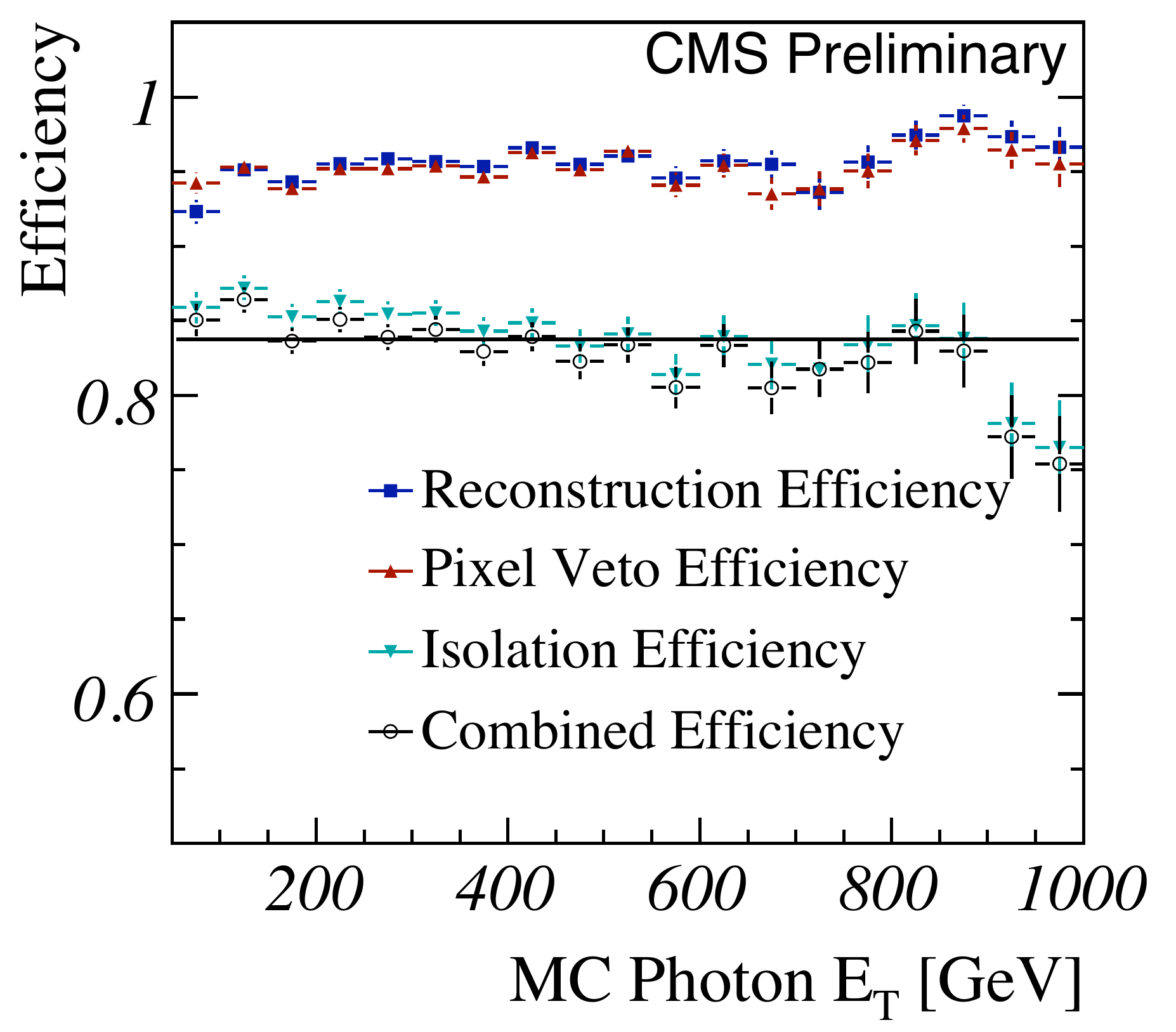}
\includegraphics[width=75 mm,height=40 mm]{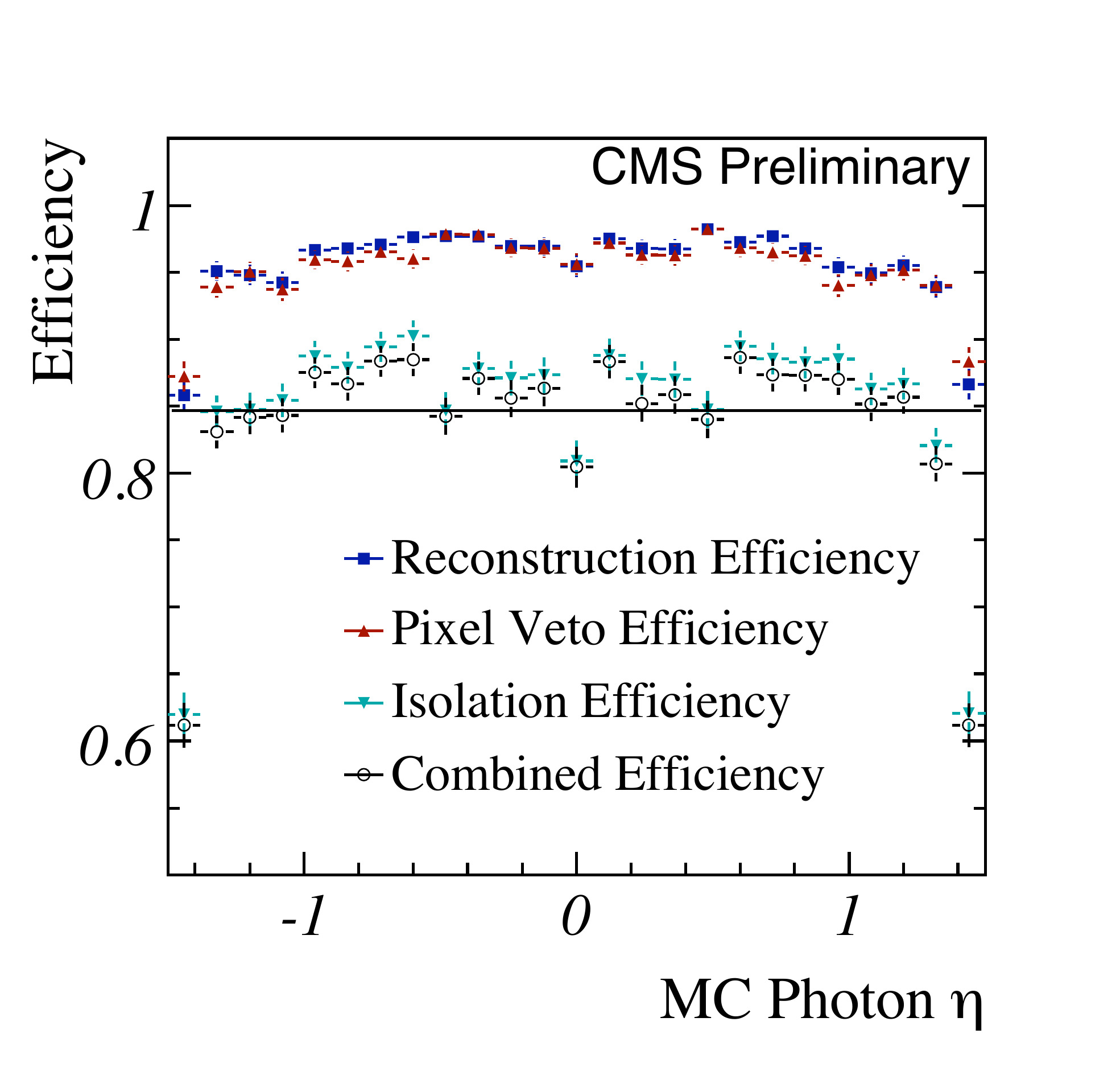}
%\resizebox{8cm}{10cm}{\includegraphics{photonid_eteff_full}}
%\resizebox{8cm}{10cm}{\includegraphics{photonid_etaeff_full}}
%\includegraphics[width=2.9in]{photonid_eteff_full}
%\includegraphics[width=2.9in]{photonid_etaeff_full}
\caption{Photon efficiency as a function of $E_{T}$ and $\eta$. Photon reconstruction, pixel detector veto, photon isolation, and the overall efficiency are shown. (The pixel detector veto and photon isolation efficiency are shown on top of the reconstruction efficiency.) In each plot, $|\eta|<1.5$.}\label{fig:photonid_eff}
\end{figure}

The efficiency of these requirements are shown in Fig~\ref{fig:photonid_eff} as a function of $E_{T}$ and $\eta$, using a high $E_{T}$ diphoton sample.  The efficiencies to reconstruct the photon candidate, pixel tracker veto, and isolation requirements are shown separately as well as combined.

%\end{figure*}

In this analysis we also also require diphotons to be $M_{\gamma\gamma}>700 GeV$.

%%%%%%%%%%%%%%%%%%%%%%%%%%%%%%%%%%
\subsection{Backgrounds}
We can categorize the backgrounds broadly into two types: irreducible and instrumental.  The instrumental backgrounds constitute the misidentification of jets and electrons as photons. Jets can fake direct photons when they fragment into a leading $\pi^0/\eta$.  Electrons can fake direct photons through tracking inefficiencies or {\it bremsstrahlung\/}.  The irreducible background is the SM production of diphotons principally through $t$-channel diagrams.  The analysis strategy is to reduce the instrumental background to be negligible compared to the irreducible direct diphoton background and then deal with the latter using kinematic cut optimization. 

\subsubsection{Jets Reconstructed as Photons}
Dijet events and direct photon production can both result in reconstructed diphoton events.  We measure the jet-faking-photon rate by using a MC dijet sample with no direct photon contamination.  The fake rate is calculated as a function of $E_{T}$ by taking the reconstructed photon $E_{T}$ spectrum and dividing it by the jet $E_{T}$ spectrum.  The ratio is fit to a polynomial curve.  We then apply this rate to jets in both dijet and direct photon events to determine the background contribution from each.
The fake rate determination depends on our ability to remove QCD direct photon contamination from dijet data (or direct diphoton contÏamination from the direct photon sample). Without such a subtraction, the fake rate will be measured higher than it should be. We studied a template fit using $\gamma\to e^+ e^-$ conversions to determine the fraction of jets and direct photons in a given sample, a method used in, e.g., Ref.~\cite{conversions}. We compare total transverse energy in the superclusters with the sum of transverse momenta of the associated tracks from photon conversions. A high $E_T/\sum p_T$ ratio is indicative of a $\pi^0/\eta \to \gamma\gamma$, as typically only one of the two photons undergoes a conversion, whereas a lower ratio indicates a converted prompt photon. We demonstrated that the method works well using dijet MC samples ~\cite{led_CMS}.

\subsubsection{Electrons Reconstructed as Photons}

We determined the electron-photon mis-identification rate, $f_{e\gamma}$, in $Z$ events using a completely data-driven approach. We reconstruct $Z$ events with $e^+ e^-$ and $e\gamma$ pairs in the $Z$  mass window.  The fake rate is then given by
\begin{equation}
f_{e\gamma} = 1-\frac{2N_{ee}}{2N_{ee}+N_{e\gamma}},
\end{equation}
where $N_{ee}$ and $N_{e\gamma}$ are the number of reconstructed $Z$'s with $ee$ and $e\gamma$ pairs, respectively.  We find that $f_{e\gamma}=(0.86 \pm 0.20)\%$, and apply this rate to high mass $e^+ e^-$ events from Drell-Yan production~\cite{led_CMS}.

\subsubsection{Diphotons}

Our strategy for this background is to assume that the MC describes the shape of the diphoton mass distribution well and then get the absolute normalization by normalizing this shape to data in a signal-depleted region ($M_{\gamma\gamma}\lesssim 500$~GeV). 

For $300$~GeV~$<M_{\gamma\gamma}<500$~GeV, the data is dominated by diphotons, although at lower mass, the direct photon and dijet contributions are non-negligible. 

Our strategy is to tighten the photon ID cuts to reduce the jet-faking-photon rate by an additional factor of 3--4.  This background reduction allows us to collect
 a relatively clean sample of direct photons at low invariant mass, which we can then use to normalize the background. 

In 100 pb$^{-1}$ of data, we can normalize the diphoton background to $15.3\pm4.9$~(stat)~$\pm1.9$~(syst) events with 100 pb$^{-1}$ of data, that is $34\%$ relative uncertainty ~\cite{led_CMS}.

\begin{figure}[!Hhtb]
\centering
\includegraphics[width=80 mm,height=45 mm]{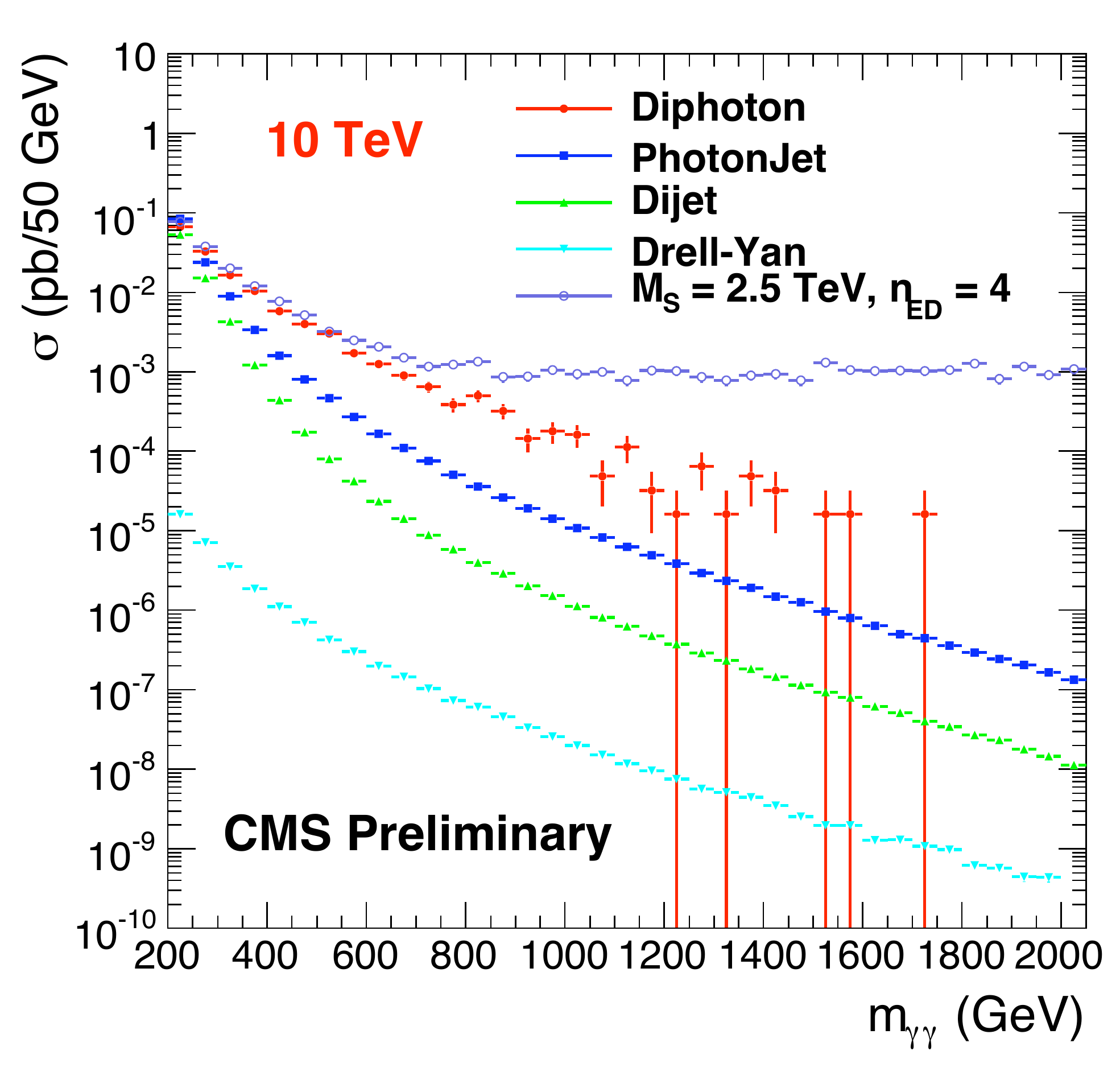}
\caption{Background and signal expectations as a function of invariant mass.  We find that the optimal point for exclusion is $M_{\gamma\gamma}>700$~GeV.  In the above plot ADD includes both SM direct diphoton production as well as the effects of large ED.\label{fig:results_invmass}}
\end{figure}

Figure~\ref{fig:results_invmass} shows the invariant mass distribution of each of the backgrounds as well as a signal distribution with the optimized $|\eta|<1.5$ requirement ~\cite{led_CMS}.

\subsection{Results}

\subsubsection{Systematics}
Early analysis is expected to be dominated by a large uncertainty on the integrated luminosity, the photon ID efficiency, and diphoton background normalization. We do not plan to do a thorough cross section analysis, so we assign a conservative 10\% uncertainty to the combined product of the diphoton efficiency and integrated luminosity. Since the dominating source of background is the diphoton production, the uncertainties on the fake background are not relevant for this analysis.  We assign a systematic uncertainty on the background of $\delta B/B = 10\% \oplus 340\%/\sqrt{\int Ldt/\mbox{\rm pb}}$, which includes a systematic uncertainty on the $K$-factor shape and the statistical uncertainty from the diphoton normalization(see details in ~\cite{led_CMS}).

\subsubsection{Limits on Large Extra Dimensions}

To establish the existence or to set limits on signal, we perform a counting experiment within the kinematic range given by the optimized cuts.  We use a Bayesian method with a flat prior chosen for the signal cross section to determine the expected limit assuming a background-only hypothesis. We further translate the cross section limits into limits on the parameters of the ADD model ~\cite{led_CMS}.

\begin{figure}[!Hhtb]
\centering
\includegraphics[width=80 mm,height=45 mm]{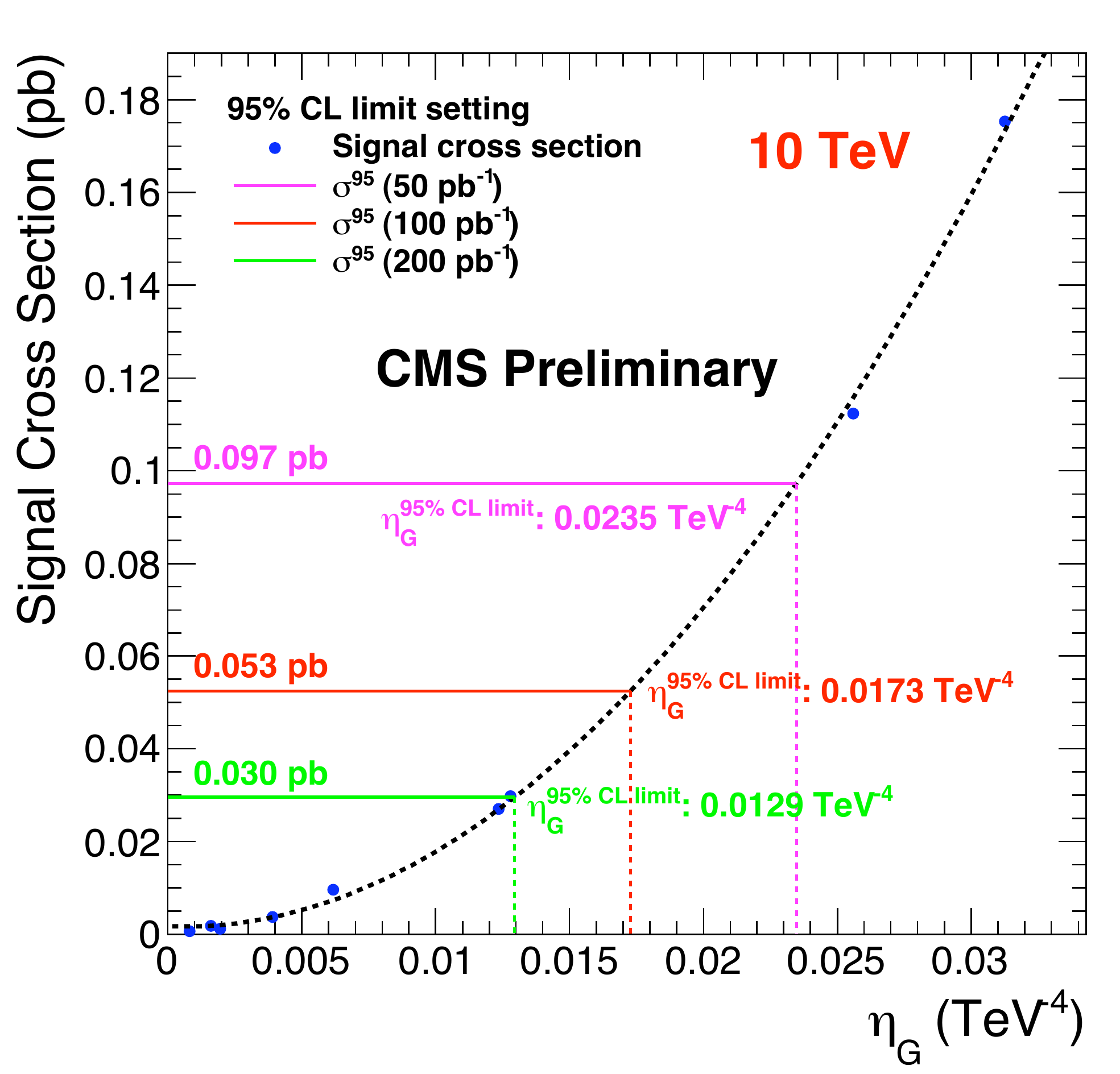}
\includegraphics[width=80 mm,height=45 mm]{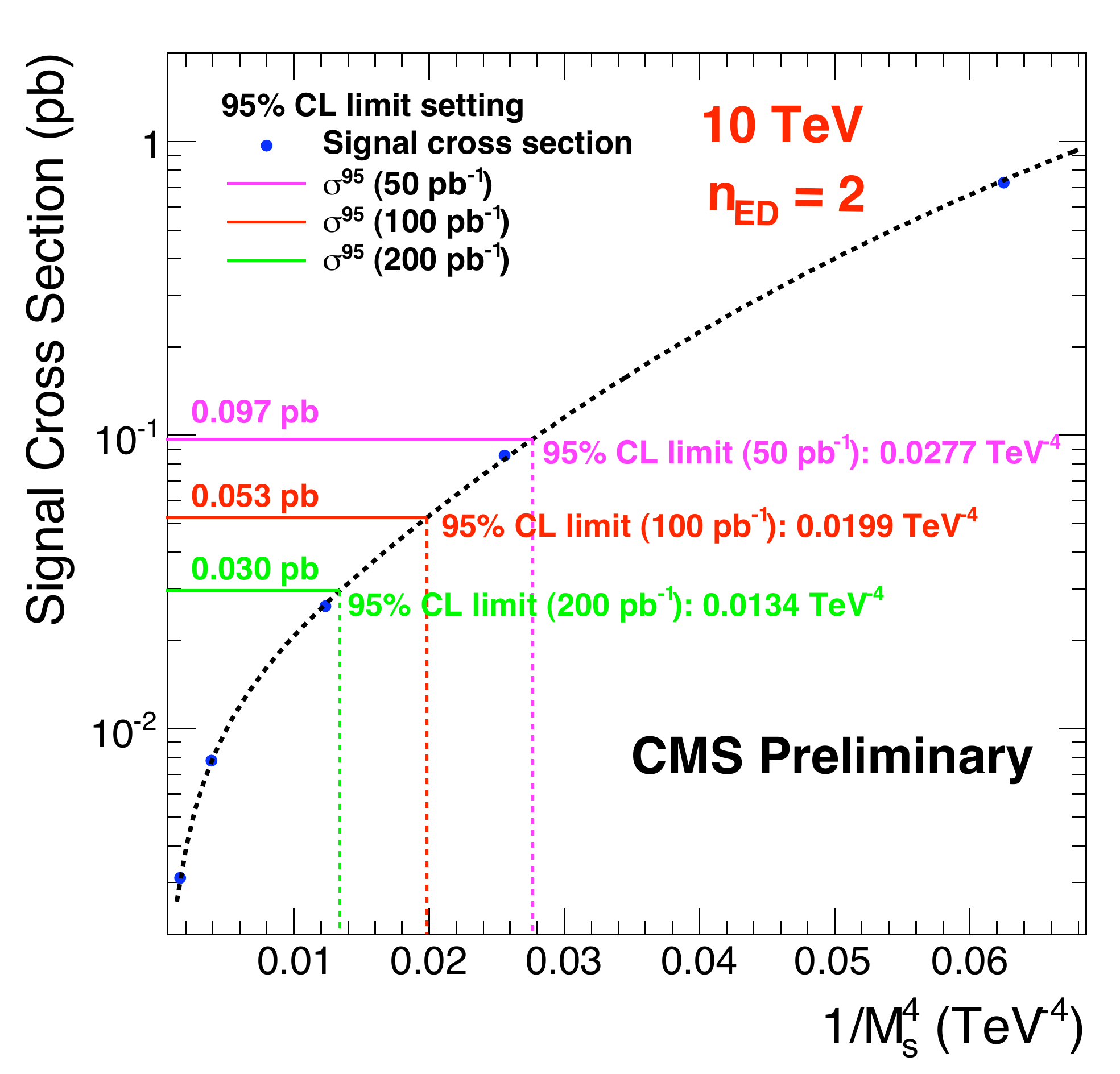}
\caption{Signal cross section parameterization as a function of the strength of the ED effect, $\eta_G$ for the $n_{ED}>2$ case (left) and as a function of $1/M_s^4$ for the $n_{ED}=2$ case (right).\label{fig:cs-limit}}
\end{figure}

The expected 95\% CL limit together with the signal cross section parameterization as a function of $\eta_G$ are shown on the left in Fig.~\ref{fig:cs-limit}. The intersection of the cross section limit with the signal cross section curve determines the upper 95\% CL limit on the parameter $\eta_G$. As seen from the plot, these limits for the 100~pb$^{-1}$ data set are equal to $\eta_G^{95} = 0.0173$~TeV$^{-4}$ and $1/M_S^4(n=2, 95\%) = 0.0199$~TeV$^{-4}$. We further translate these limits into the lower limit on the fundamental Planck scale for various numbers of extra dimensions $n_{\rm ED}$, as shown in Table~\ref{table:limits}. This is calculated trivially for $n_{\rm ED} = 2$ and for $n_{\rm ED}>2$ by using Eq.~\ref{eq:HLZ}. In addition, we quote 95\% exclusion for integrated luminosities of 50 and 200 pb$^{-1}$. These limits are shown in Fig.~\ref{fig:cs-limit}, as well as in Table~\ref{table:limits}.

\begin{table}
\centering
\begin{tabular}{cccc}
\hline
 $n_{ED}$        & \multicolumn{3}{c}{95\% CL Limit on $M_S$} \\
 &  50~pb$^{-1}$ & 100~pb$^{-1}$ & 200~pb$^{-1}$        \\
\hline
 2              &       2.5 TeV & 2.7 TeV & 2.9 TeV \\
 3              &       3.0 TeV & 3.3 TeV & 3.5 TeV \\
 4              &       2.6 TeV & 2.8 TeV & 3.0 TeV \\
 5              &       2.3 TeV & 2.5 TeV & 2.7 TeV \\
 6              &       2.1 TeV & 2.3 TeV & 2.5 TeV \\
 7              &       2.0 TeV & 2.2 TeV & 2.4 TeV \\
\hline
\end{tabular}
\caption{Table of $95\%$ CL limits on $M_S$ as a function of the
number of ED for three characteristic integrated luminosities expected
to be reached in 2010.\label{table:limits}}
\end{table}

Note that the limits on the fundamental Planck scale expected from this search with $\sim 100$~pb$^{-1}$ of data are approximately twice as high as the best sensitivity achieved so far (at the Tevatron)~\cite{D0diphoton} and represent a significant extension in the previously unexplored region of the parameter space of the ADD model.

\subsubsection{Discovery Potential for Extra Dimensions}

\begin{table}[tb]
\centering
\begin{tabular}{ccc}
\hline
 ED Parameters  & $\int Ldt$ for 3$\sigma$      &
$\int Ldt$ for 5$\sigma$     \\
\hline
$M_s=2$~TeV, $n_{ED}=2$ & $\sim12$~pb$^{-1}$ & $\sim20$~pb$^{-1}$\\
$M_s=2$~TeV, $n_{ED}=4$ & $\sim7$~pb$^{-1}$ & $\sim11$~pb$^{-1}$\\
$M_s=2$~TeV, $n_{ED}=6$ & $\sim32$~pb$^{-1}$ & $\sim72$~pb$^{-1}$ \\
$M_s=2.5$~TeV, $n_{ED}=2$ & $\sim62$~pb$^{-1}$ & $\sim162$~pb$^{-1}$ \\
$M_s=2.5$~TeV, $n_{ED}=4$ & $\sim51$~pb$^{-1}$ & $\sim129$~pb$^{-1}$ \\
$M_s=2.5$~TeV, $n_{ED}=6$ & $\sim342$~pb$^{-1}$ & $\sim$914~pb$^{-1}$ \\
$M_s=3$~TeV, $n_{ED}=2$ & $\sim314$~pb$^{-1}$ & $\sim$846~pb$^{-1}$\\
$M_s=3$~TeV, $n_{ED}=4$ & $\sim387$~pb$^{-1}$ & $\sim$1050~pb$^{-1}$ \\
\hline
\end{tabular}
\caption{Luminosity needed for observation or discovery given the
$M_S$ and $n_{ED}$ parameters.
\label{table:discovery}}
\end{table}

To estimate the discovery potential, we calculate the Poisson probability for the background to fluctuate to or above the number of events $n$ observed in the counting window.  We further convert the $p$-value into the Gaussian significance of a one-sided fluctuation, represented as a number of standard deviations, $\sigma$.  However, we want to ensure that the discovery is not claimed based on just one event observed. Therefore, we also control the number of expected events, $S+B$ and require this number to be at least 3 to claim $3\sigma$ evidence and 5 to claim a $5\sigma$ discovery ~\cite{led_CMS}. (This is not a mathematically strict, but nevertheless a popular choice in the literature, which, for one, ensures that an observation luminosity always exceeds the 95\% CL limit luminosity.)

\begin{figure}[tb]
\centering
\includegraphics[width=80 mm,height=45 mm]{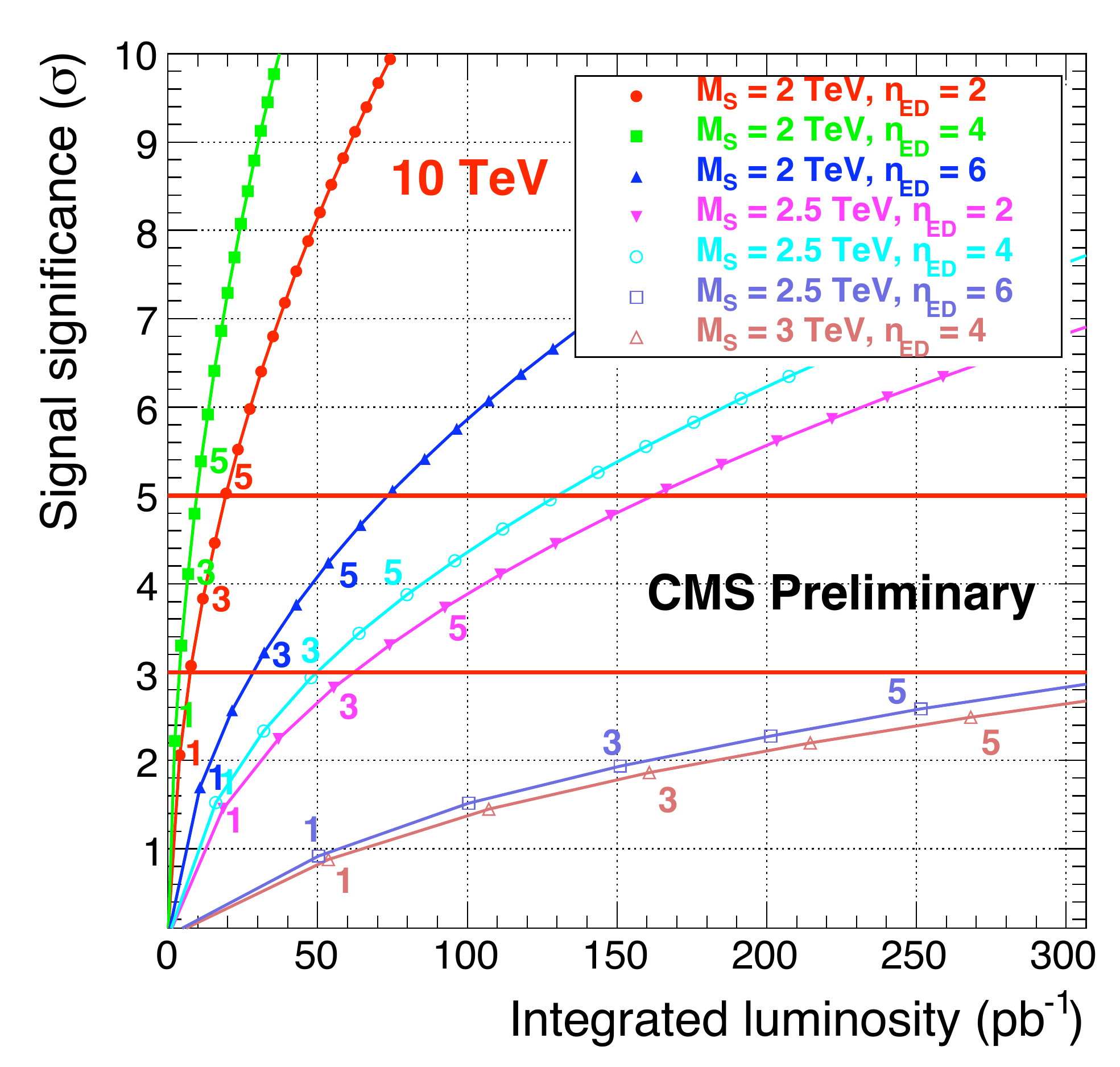}
\caption{Luminosity required for signal discovery for different values of $M_S$ and number of ED.  Shown on the $y$-axis is the corresponding $p$-value (in one-sided Gaussian $\sigma$'s). Points on the lines indicate consecutive integer number of signal events; points corresponding to 1, 3, and 5 events are labeled accordingly. Requiring at least 3 events for observation or 5 events for discovery ensures that discovery cannot be claimed based on a single, golden event.\label{fig:discovery}}
\end{figure}

The results for our counting experiment are shown in Fig.~\ref{fig:discovery} and also listed in Table~\ref{table:discovery}. As one can see, a $5\sigma$ discovery for the case of $M_S = 2$~TeV and $n_{\rm ED} \le 6$ is possible with less than 75 pb$^{-1}$ of data With 130~pb$^{-1}$ the discovery is possible up to $M_S = 2.5$~TeV and $n_{\rm ED} \le 4$. For $M_S \ge 3$~TeV the first LHC run is not expected to be sufficient to claim either the discovery, or even a $3\sigma$ evidence for the signal. However, even with 100~pb$^{-1}$ of data a considerable region of previously unexplored parameter space can be probed with the discovery sensitivity.
    
\section{Warped Extra Dimensions}

The Randall-Sundrum (RS) scenario suggests that the observed hierarchy is created by a warped extra dimension~\cite{rs}.

RS models posit that our universe is described by a higher-dimensional warped geometry.  The warping of the extra dimension causes the energy scale of one end of the extra dimension to be much larger than at the other end.
%%%
In the simplest RS models, our universe is represented as a 5-dimensional space, and all particles, except for the graviton, are localized on (3 + 1)-dimensional brane(s).  

We consider RS-1 model, which have a finite size for the extra dimension, with two branes, one at each end. The observed hierarchy is generated by a geometrical exponential factor; gravity originates on the Planck brane and the graviton wave function is exponentially suppressed as we move along the extra dimension away from the Planck brane.  
  
Gravitons appear as a tower of Kaluza-Klein excitations with masses and widths determined by the parameters of the RS-1 model:  
the mass of the first graviton excitation mode $M_1$, or $M_{G}$, and the dimensionless coupling parameter $\tilde{k} = k/\bar{M}_{Pl}$. Precision electroweak data require that $\tilde{k}>0.01$, while the requirement that the model remains perturbative constrains $\tilde{k}<0.1$.

This analysis utilizes the results of the search  for graviton decays in the diphoton channel of the ADD model of large ED~\cite{led_CMS}. The strategy for this diphoton resonance search is to use the results from the non-resonant large ED search to extract limits on and probe the possibility discovery of the RS warped extra dimension model~\cite{RS_CMS}.  This approach is slightly less sensitive than a dedicated analysis for the search for a diphoton resonance.  However, since the backgrounds are small, the limits from such an approach will not be affected significantly.

\subsection{Signal and Event Selection}
The signal for our search consists of two high energy photons, arising from the decay of an RS-1 graviton. The signal samples span $M_{1}=750, 1000, 1250$, and $1500$ GeV with $\tilde{k}=0.01$. 

The dominant background for the diphoton decay of the RS-1 graviton is SM diphoton production and the instrumental backgrounds are negligible. A detailed description of background Monte Carlo samples and the procedures used for estimating backgrounds with data-driven methods can be found in the large ED note~\cite{led_CMS}.  

For this analysis, we apply the same event selection requirements as used in the large ED analysis, as follows:  Photon $p_T > 50$ GeV , Photon $|\eta| < 1.5$, and $M_{\gamma\gamma} > 700$ GeV.

\begin{figure}[tb!]
  \begin{center}
    \includegraphics[width=80 mm,height=45 mm]{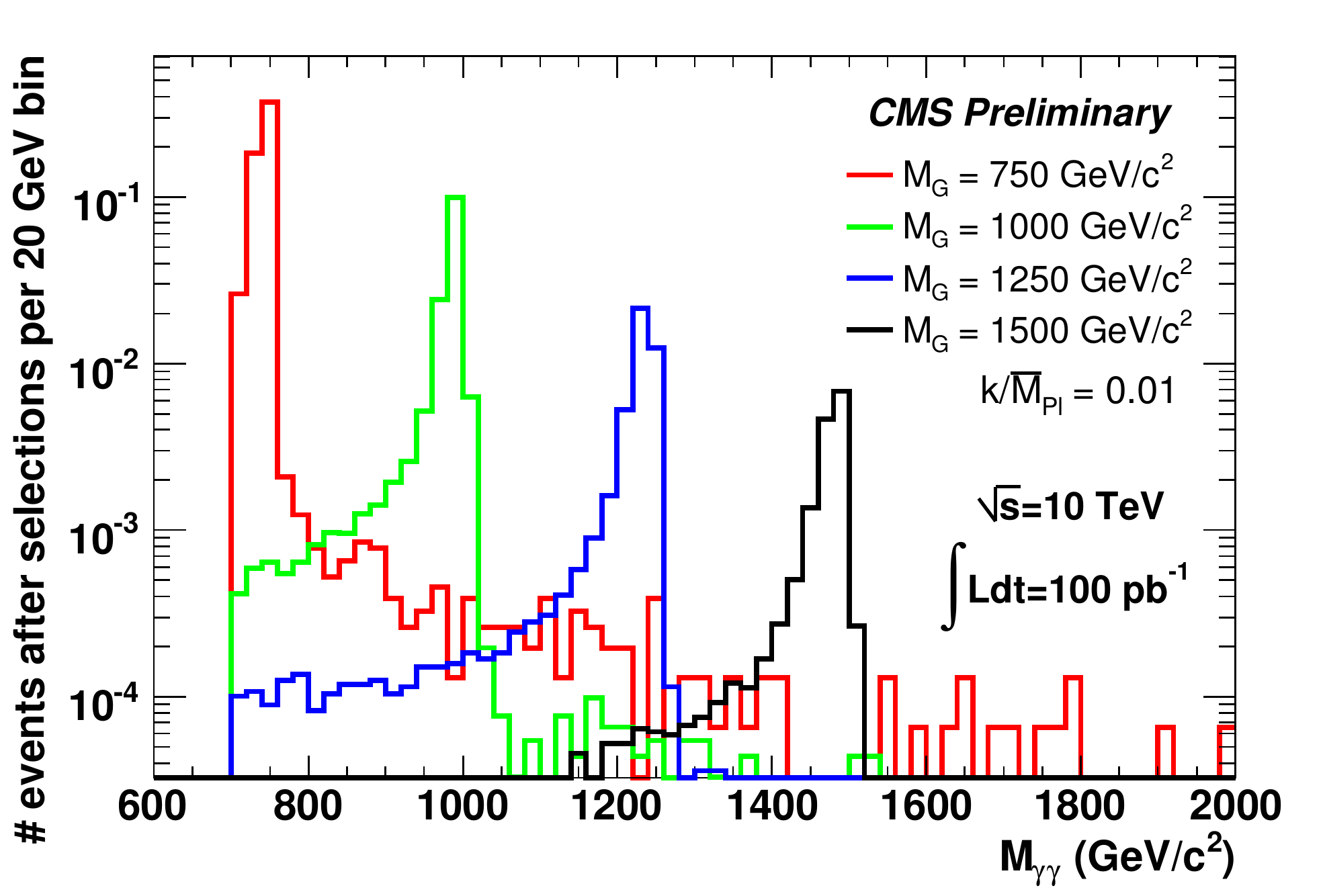}
    \caption{Diphoton invariant mass spectra after selection is applied, scaled to 100 pb$^{-1}$ for $M_{1}=750, 1000, 1250$, and $1500$ GeV, $\tilde{k}=0.01$ samples.}
    \label{fig:rsmass}
  \end{center}
\end{figure}

The plots in Figure~\ref{fig:rsmass} show the diphoton invariant mass spectra for the signal samples, after the selection is applied and scaled to 100 pb$^{-1}$.

In this analysis, we consider the signal acceptance (after cuts are applied, as measured for MC truth particles) and the photon reconstruction and identification efficiencies separately. While the signal acceptance is between $50\% - 67\%$ the efficiency for reconstructing and identifying photons is between $63\% - 59\%$, depending on the mass (see the details in ~\cite{RS_CMS}).
The photon reconstruction and identification requirements are described in detail in the large ED analysis note~\cite{led_CMS}.  

 It should be noted that the diphoton efficiency measured by the large ED analysis was $72\pm 7 \%$, which differs from what is measured for the RS analysis; this difference is taken into account in the calculation of the limits and discovery reach, as described in the following sections.  

\subsection{Results}

\subsubsection {Limits on RS Model Parameters}\label{sec:limits}

After calculating the acceptance after selection for our RS-1 $G_{KK}$ samples, we extrapolate these rates to limits on the model parameters: the graviton mass $M_1$ and the coupling parameter $\tilde{k}$.  
Here, we take advantage of the results of the non-resonant large ED graviton search. The details for the extrapolation of the large ED limit can be found in the ~\cite{RS_CMS}.

\begin{figure}[bt!]
  \begin{center}
    \includegraphics[width=80 mm,height=45 mm]{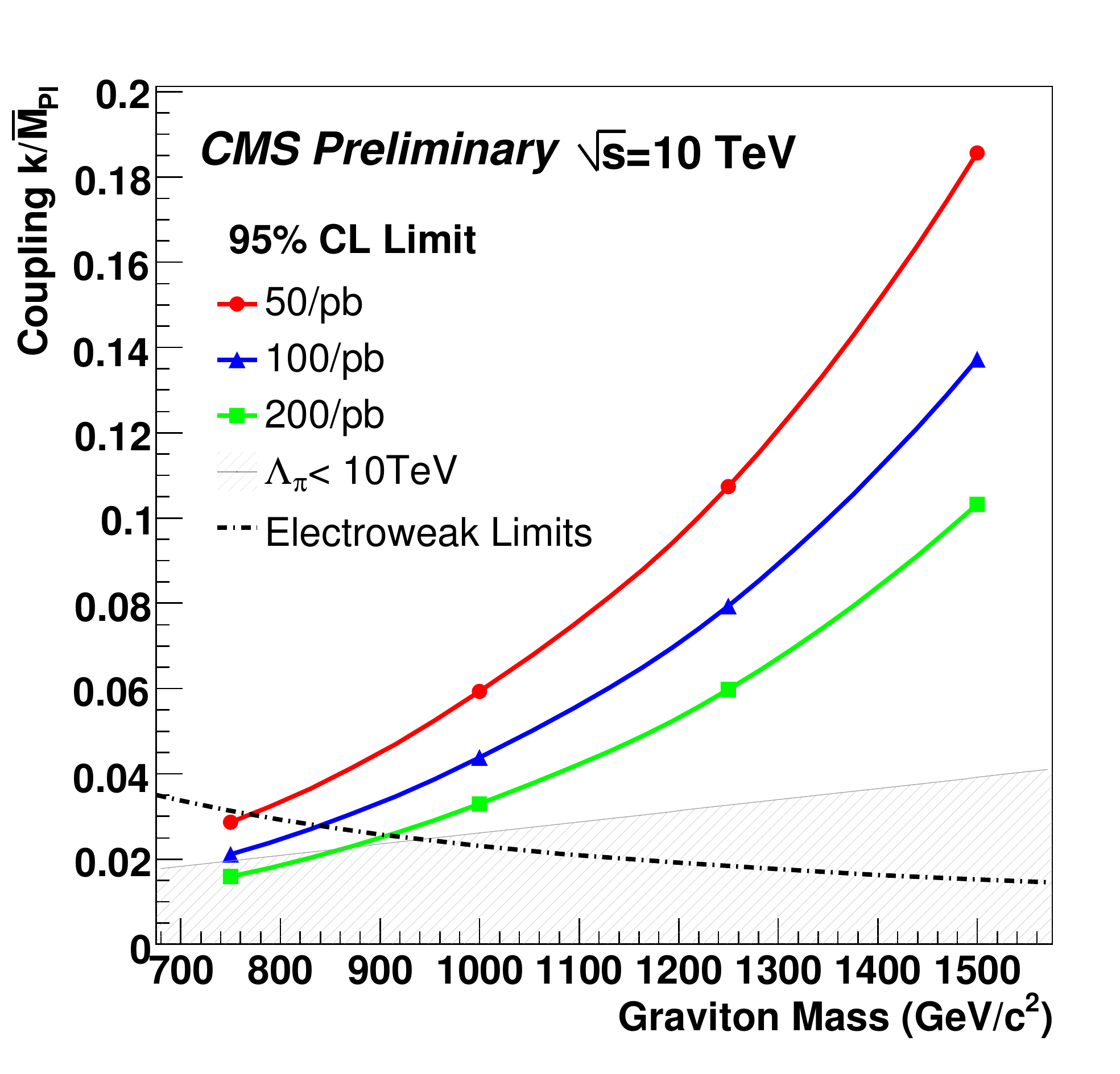}
    \caption{Limit on RS parameters $(M_{1},\tilde{k})$, extrapolated from the results of the large ED diphoton search for 100/pb.  The area to the left of the curves is excluded.  The gray shaded region shows the area excluded for $\Lambda_{\pi}<10$ TeV.  The area below the dash-dotted line is excluded by precision electroweak data~\cite{ewk}. }
    \label{fig:limit}
  \end{center}
\end{figure}

 Figure~\ref{fig:limit}, which shows the extrapolated limit in the $(M_1,\tilde{k})$ plane.  We see from this figure that with 50 pb$^{-1}$, we can place a $95\%$ lower limit on a graviton of mass up to roughly 1.2 TeV with $\tilde{k}$=0.1.  Likewise, with 100 pb$^{-1}$, we can place a lower limit on a graviton mass up to 1.35 TeV with $\tilde{k}$=0.1. We extend the sensitivity beyond the reach obtained at the Tevatron ~\cite{cdf, d0}.

\subsubsection{Discovery Potential for RS Warped Extra Dimension}

We also consider the signal discovery potential for the RS-1 warped extra dimension model and calculate the estimated integrated luminosity needed to establish a signal. To estimate the discovery potential we used the same method as used in the large ED ~\cite{led_CMS}.  It should be noted that the photon efficiency has been taken into account in this calculation, as described in Section ~\ref{sec:limits}.

\begin{figure}[hbtp]
  \begin{center}
     \includegraphics[width=80 mm,height=45 mm]{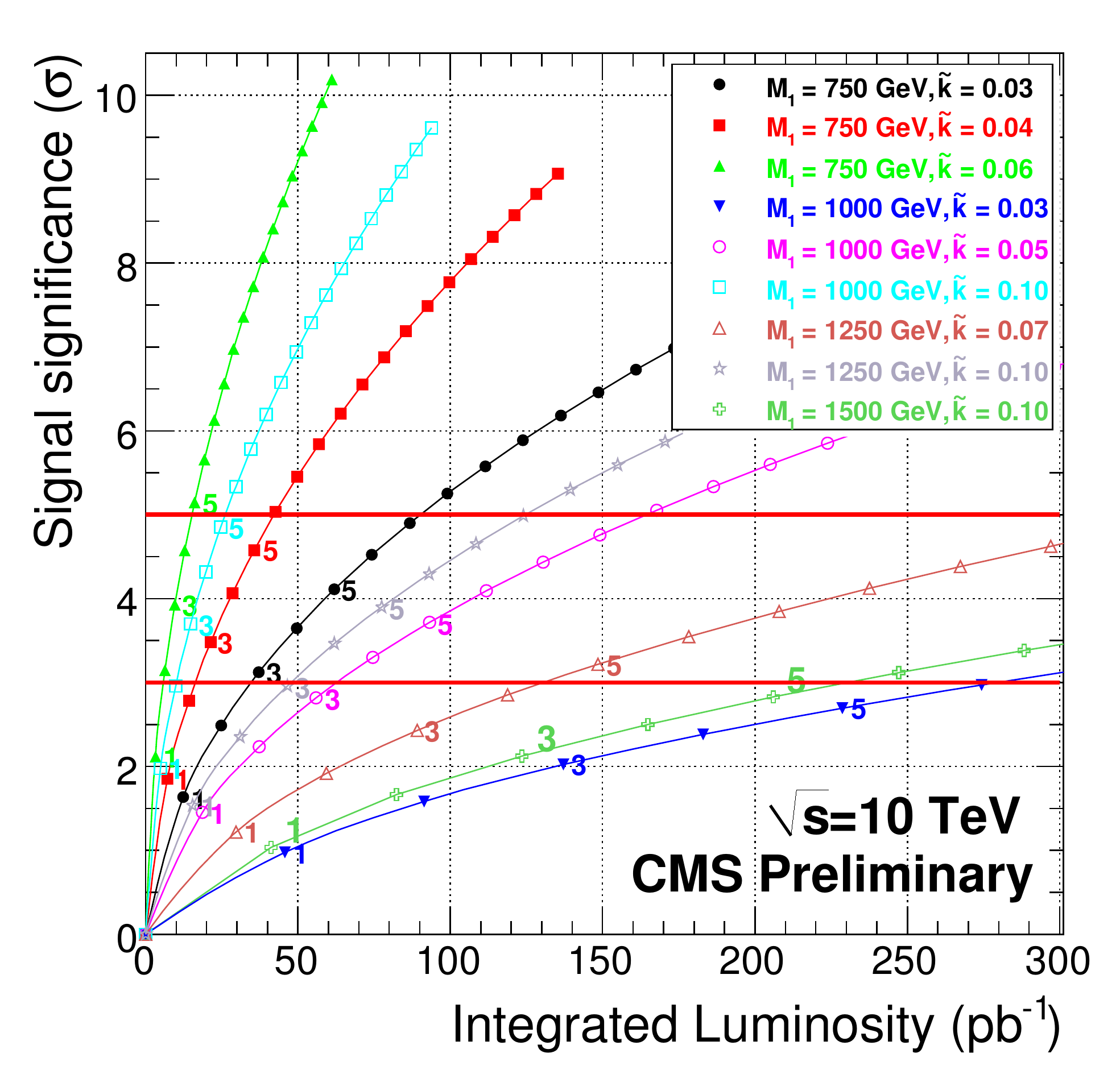}
    \caption{Luminosity required for signal discovery for various values of the RS-1 model parameters $M_{1}$ and $\tilde{k}$.  The $y$-axis shows the corresponding $p$ value (in one-sided Gaussian $\sigma$'s). Requiring at least 3 events for observation and 5 for discovery ensures that the discovery cannot be claimed on a single event.}
    \label{fig:disc}
  \end{center}
\end{figure}

The results are shown in Fig~\ref{fig:disc}.  
%We can see that a $5\sigma$ discovery
We can see that with as little as 30 pb$^{-1}$ we can claim a $5\sigma$ discovery for a 1 TeV mass graviton with $\tilde{k}=0.1$.  
%With $60\pbinv$ we can claim discovery for a 1.25\TeVcc\ graviton with $\tilde{k}=0.1$.  
With 100 pb$^{-1}$, we can claim $5\sigma$ discovery for a 750 GeV RS graviton with $\tilde{k}=0.03$, and with 130 pb$^{-1}$ we can claim a discovery for a 1.25 TeV RS graviton with $\tilde{k}=0.1$.
We see that, even with a small sample of early LHC data, we can cover a large range of the parameter space, particularly if the coupling is strong $\tilde{k}\sim0.1$ ~\cite{RS_CMS}.

\section{Conclusion}

To summarize, we performed a MC study of the sensitivity of the CMS experiment to models with extra spatial dimensions~\cite{add} in the diphoton final state in the 2009--2010 LHC run at the center-of-mass energy of 10 TeV ~\cite{led_CMS, RS_CMS}. We developed identification cuts essential for photons at large transverse momenta, typical of such a signal, and showed that instrumental backgrounds can be kept under control. In the absence of an excess over the dominant SM direct diphoton background in a 100~pb$^{-1}$ data set, we set an upper limit on the parameter $\eta_G$ of the ADD models of 0.0173 TeV$^{-4}$ at the 95\% CL, which translates in the lower limits on the effective Planck scale of $M_S > 2.8$~TeV for $n_{\rm ED} = 4$, which is twice as stringent as the best limits to date coming from the Tevatron~\cite{D0diphoton}. In the presence of the signal, it can be observed at a 5 standard deviation level up to $M_S = 2.5$~TeV and $n_{\rm ED} = 4$ with $\approx 130$~pb$^{-1}$ of data. We also investigate the feasibility of discovering a high  mass diphoton resonance \cite{rs}  with the CMS detector.  We place limits on the RS-1 parameters $(M_{1}, \tilde{k})$ and probe the discovery potential of the RS-1 model, utilizing the results of the ADD graviton search in the diphoton channel. 
With 100 pb$^{-1}$, we can place a $95\%$ CL limit on a graviton mass up to 1.35 TeV with $\tilde{k}$=0.1.  
Likewise, with 100 pb$^{-1}$, we can claim $5\sigma$ discovery for a 750 GeV RS graviton with $\tilde{k}=0.03$.
%Likewise, with $100\pbinv$, we can claim $5\sigma$ discovery for a 1.0 \TeVcc\ RS graviton with $\tilde{k}=0.1$.
Even with a small sample of early LHC data, we can cover a large range of the parameter space, particularly if the coupling of the model is strong $\tilde{k}\sim0.1$.

%%%%%%%%%%%%%%%%%%%%%%%%%%%%%%%%%%
\begin{acknowledgements}
I would like to thank my supervisor at Brown, Greg Landsberg, for giving me the opportunity to work on the Extra Dimensions analysis, and present it at the DPF'09 Conference. I also thank my colleagues D. Nguyen, J.P. Chou at Brown,  M. Gataullin, V. Litvin, T. Orimoto, and at Caltech for the fruitful colloboration. Special thanks go to John Strologas. I am grateful to the entire CMS Collaboration, the technical and administrative staff at CERN. And, my deepest love goes to O\u{g}uzhan...

\end{acknowledgements}

%%%%%%%%%%%%%%%%%%%%%%%%%%%%%%%%%%


\bigskip % extra skip inserted

\begin{thebibliography}{9}   % Use for  1-9  references
%\begin{thebibliography}{99} % Use for 10-99 references

\bibitem {add} N. Arkani-Hamed, S. Dimopoulos, G.Dvali,
  Phys. Lett. {\bf B249} (1998) 263\\
 I. Antoniadis,  N. Arkani-Hamed, S. Dimopoulos, G.Dvali,
  Phys. Lett. {\bf B436} (1998) 257

\bibitem{rs} L. Randall, R. Sundrum, Phys. Rev. Lett. {\bf 83} (1999)
  3370 and {\it ibid} (1999) 4690.

\bibitem{intro} I. Antoniadis, Phys. Lett. {\bf B246} (1990) 377 \\
 I. Antoniadis, C. Munoz, M.Quiros, Nucl. Phys. {\bf B397} (1993) 525 \\
 I. Antoniadis, K. Benalki, Phys. Lett. {\bf B236} (1994) 69

\bibitem{HLZ}
    T. Han, J. Lykken, and R.J. Zhang, Phys. Rev. D {\bf 59}, 105006 (1999).

\bibitem{GRW}
    G. Giudice, R. Rattazzi, and J. Wells, Nucl. Phys. B {\bf 544}, 3 (1999).

\bibitem{Hewett}
    J. Hewett, Phys. Rev. Lett. {\bf 82}, 4765 (1999).

\bibitem{conversions} M.~D'Alfonso and others,CMS Physics Analysis Summary (PAS),\href{http://cms-physics.web.cern.ch/cms-physics/public/SUS-08-002-pas.pdf}.

\bibitem{Peskin} E.A.~Mirabelli and M.~Perelstein and M.E.~Peskin, Phys.\ Rev.\ Lett. {\bf 82}, 2236 (1999).


\bibitem{D0diphoton} V. M. Abazov and others, Phys. Rev. Lett., {\bf 102}, 051601 (2009)

\bibitem{cdf}
CDF Collaboration,  T.~Aaltonen {\it et al.},
%{``Search for a high-mass diphoton state and limits on Randall-Sundrum gravitons at CDF,''} 
Phys.\ Rev.\ Lett.\  {\bf 99}, 171801 (2007).
% [arXiv:0707.2294 [hep-ex]].
%%CITATION = PRLTA,99,171801;%%

\bibitem{d0}
D0 Collaboration, V.~M.~Abazov {\it et al.}, 
%``Search for Randall-Sundrum gravitons with 1 $fb^{-1}$ of data from $p\bar{p}$ collisions at $\sqrt{s}$ = 1.96-TeV,'' 
Phys.\ Rev.\ Lett.\  {\bf 100} (2008) 091802.

\bibitem{led_CMS}
CMS Collaboration,
``Search for Large Extra Dimensions in the Diphoton Final State'', CMS PAS EXO-09-004, 2009, 10pp. 
http://cms-physics.web.cern.ch/cms-physics/public/EXO-09-004-pas.pdf

\bibitem{RS_CMS}
CMS Collaboration,
``Search for Randall-Sundrum Gravitons in the Diphoton Final State'', CMS PAS EXO-09-009, 2009, 10pp. http://cms-physics.web.cern.ch/cms-physics/public/EXO-09-009-pas.pdf

\bibitem{ewk}
H. Davoudiasl, J.L. Hewett, and T.G. Rizzo, 
Phys.\ Rev.\ D{\bf63}, 075004 (2001).

\end{thebibliography}

\end{document}